\documentstyle [12pt]{article}

\title {\LARGE{Monte Carlo Calculation of Phase Shift}\\
\LARGE{in Four Dimensional O(4) $\phi^4$ Theory}}
\author{
{}~\\~\\
\large{\sc Jun Nishimura}
\thanks{e-mail address : {\tt nishi@tkyvax.phys.s.u-tokyo.ac.jp} (internet)}\\
{}~\\
\normalsize{{\it Department~of~Physics,~University~of~Tokyo}} \\
\normalsize{{\it Bunkyo-ku,~Tokyo~113,~Japan}}\\
{}~\\~\\~
}

\date{26 July 1992}

\begin{document}
\maketitle
\vspace{10mm}

\newcommand {\beq}{\begin{equation}}
\newcommand {\eeq}{\end{equation}}
\newcommand {\beqa}{\begin{eqnarray}}
\newcommand {\eeqa}{\end{eqnarray}}
\newcommand {\beqan}{\begin{eqnarray*}}
\newcommand {\eeqan}{\end{eqnarray*}}
\newcommand {\n}{\nonumber \\}
\newcommand {\cleqn}{\setcounter{equation}{0}}
\newcommand {\eq}[1]{eq.~(\ref{#1})}
\newcommand {\eqs}[1]{eqs.~(\ref{#1})}
\newcommand {\Label}[1]{\label{#1}}
\newcommand {\Bibitem}[1]{\bibitem{#1}}
\newcommand {\Romannumeral}[1]{\uppercase\expandafter{\romannumeral#1}}
\newcommand {\ds}{\displaystyle}
\newcommand \bra[1]{\left< {#1} \,\right\vert}
\newcommand \ket[1]{\left\vert\, {#1} \, \right>}
\newcommand \braket[2]{\hbox{$\left< {#1} \,\vrule\, {#2} \right>$}}
\newcommand \vev[1]{\hbox{$\left< \,{#1} \, \right>$}}

\newcommand {\ee}{\mbox{e}}
\newcommand {\rb}{\mbox{{\bf r}}}
\newcommand {\nb}{\mbox{{\bf n}}}
\newcommand {\xb}{\mbox{{\bf x}}}
\newcommand {\tb}{\mbox{{\bf t}}}
\newcommand {\pbb}{\mbox{{\bf p}}}
\newcommand {\kb}{\mbox{{\bf k}}}
\newcommand {\ob}{\mbox{{\bf o}}}
\newcommand {\ub}{\mbox{{\bf u}}}
\newcommand {\vb}{\mbox{{\bf v}}}
\newcommand {\dd}{\mbox{d}}
\newcommand {\rprn}{\mbox{]}}
\newcommand {\lprn}{\mbox{[}}
\newcommand {\rg}{\rangle}
\newcommand {\lgg}{\langle}
\newcommand {\del}{\partial}

\begin{abstract}
\noindent
The phase shift of the O(4) symmetric $\phi^4$ theory in the
symmetric phase is calculated numerically using the relation between phase
shift
and energy levels of two-particle states recently derived by L\"{u}scher.
The results agree with the prediction of perturbation theory.
A practical difficulty of the method for a reliable extraction of the phase
shift for large momenta due to the necessity of a precise determination of
excited two-particle energy levels is pointed out.
 \end{abstract}

\newpage

Monte Carlo simulation techniques enable us to
study various non-perturb-ative phenomena of field theories defined on a
space-time lattice. Because of the Euclidean formalism of lattice field
theories, however, the method does not allow a direct evaluation of
physical quantities characterizing scattering processes such as phase shift  .

It has been known for some time\cite{Hamber1,Luescher0,Luescher2} that
finite-size effects in energy levels are closely related to scattering
amplitudes.
Indeed, asymptotic behavior of energy levels of two-particle states
for large volume can be written in terms of scattering length
\cite{Luescher2}
(see Refs. \cite{Montvay,Frick,Guagnelli,Sharpe} for its applications).
Recently this asymptotic formula has been generalized to an exact
relation between
energy levels of two-particle states in a finite box and phase shift
\cite{Luescher1}.
Since two-particle
energy levels are calculable through standard Monte Carlo techniques, this
relation opened a possibility  of extracting phase shift through numerical
simulations.  Of particular interest is that the relation may allow a
determination of resonance parameters in QCD\cite{Luescher3}.

The corresponding relation in two dimension has been used to numerically
extract the phase shift of the O(3) non-linear $\sigma$
model\cite{LuescherWolff}  and a coupled Ising system\cite{Lang}
and  a good agreement has been found between the numerical results
and analytic predictions.  These results encourage us to  examine the practical
applicability of the method for realistic field theories in four dimensions.
In
this article we report on our attempt to extract the phase shift of the
four-dimensional O(4) symmetric $\phi^4$ theory in the symmetric phase.
This model provides a good testing ground of the method since the triviality of
the theory, which is  confirmed in many ways\cite{Hasenfratz0} though no exact
proof exists,  enables us to check the results against perturbative
calculations.

Consider a system of two identical particles in a cubic box of a size $L^3$,
whose states are classified by the irreducible
representations of the cubic group SO(3,${\bf Z}$).  Let
$W_j~(j=0,1,2,\cdots;~W_0<W_1<\cdots)$ be the energy levels of the two-particle
states in an irreducible representation of  SO(3,${\bf Z}$).  Let $m$ be the
mass of the particle and define the momentum $k_j$ corresponding to  $W_j$ by
\beq
   W_j=2\sqrt{m^2+k_j^2}.
\label{eq:dispersion}
\eeq
L\"{u}scher's formula\cite{Luescher1}
relates $k_j$ to the set of phase shifts $\delta_l(k_j)$  having the angular
momentum $l$ not excluded by the symmetry of the two-particle state.  For the
states in the $A_1^+$ representation the relevant phase shifts are
$\delta_0$, $\delta_4$, $\delta_6$, $\cdots$.  At low energies the s-wave phase
shift $\delta_0$ dominates.  Neglecting the phase shifts $\delta_l$ with $l\geq
4$ the formula takes the form\cite{Luescher1}   \beq
   -\delta_0 (k_j)=\phi \left( \frac{k_jL}{2\pi} \right) - j\pi,
\label{eq:luescher}
\eeq
where $\phi(q)$ is given by
\beq
   e^{-2i\phi(q)}={{\cal Z}_{00}(1;q^2)+i\pi^{3/2}q\over
                   {\cal Z}_{00}(1;q^2)-i\pi^{3/2}q},~~~~~\phi(0)=0,
\eeq
with ${\cal Z}_{00}(1;q^2)$ defined by an analytic continuation of
\beq
   {\cal Z}_{00}(s;q^2)={1\over \sqrt{4\pi}}\sum_{{\bf n}\in {\bf Z}^3}{1\over
({\bf n}^2-q^2)^s}.
\eeq
In a numerical simulation of a given size $L$, the formula (\ref{eq:luescher})
determines the phase shift only at a discrete set of momenta $k_0,k_1,\cdots$
corresponding to the energy levels $W_0, W_1, \cdots$.  The momenta,
however, can be shifted by using a different lattice size.  Combining the
results for various lattice sizes one can obtain the full momentum dependence
of
the phase shift.

We apply the method above to the $\phi^4$ theory in four dimensions
with  O(4) symmetry, employing the standard lattice action given by
\beq
    S=-2\kappa\sum_{n \mu}\sum_{\alpha=1}^{4}\phi_n^{\alpha}
      \phi_{n+\hat{\mu}}^{\alpha},~~~~~
\sum_{\alpha=1}^{4}\phi_n^{\alpha}\phi_n^{\alpha}=1.
\eeq
The system undergoes a second-order phase transition at $\kappa=\kappa_c\equiv
0.30411(6)$, above which the O(4) symmetry is spontaneously broken.
The simulations are made at $\kappa=0.297$ in the symmetric phase
on an $L^3\times 16$ lattice with $L=8$ and 12.
We carried out $4\times 10^6$ sweeps of the heat bath algorithm for each
lattice size,
measuring observables at every 10 sweeps.

We extract single-particle energies
from the exponential decay of the propagator
\beq
\sum_{\alpha} \langle \phi_{\bf p}^{\alpha}(t) ^{\ast}
\phi_{\bf p}^{\alpha}(0) \rangle \longrightarrow {\rm const.}
\ee^{-E({\bf p})t}
\eeq
with $\phi_{\bf p}^{\alpha}(t)$ the projection of $\phi_n^{\alpha}$ to
the spatial momentum
${\bf p}$ at a time slice $t$.  The exponential fits are
made over the range $t=3-6$ for the spatial size $L=8$ and $t=4-6$ for $L=12$.
The results for the momenta ${\bf p}=(0,0,0), (1,0,0), (1,1,0)$ and $(1,1,1)$
(in units of $2\pi/L$) are given in Table 1.  The errors are estimated by the
jackknife method with the bin size of $10^6$ sweeps. Our values for the mass
$m=E({\bf 0})$ are consistent with those of Ref. \cite{Frick} for the same
choice of
$\kappa$ and $L$.

The O(4)-scalar two-particle operator for a center-of-mass momentum {\bf
p} is given by
\beq
  {\cal O}_{\mbox{{\scriptsize\bf p}}}(t)=
\sum_{\alpha}\sum_{R} \phi_{ R \mbox{{\scriptsize\bf p}}}^{\alpha}(t)
      \phi_{ -R \mbox{{\scriptsize\bf p}}} ^{\alpha}(t)
\eeq
where the summation over cubic rotation $R$ ensures the projection onto the
$A_1^+$-sector.
The corresponding two-point function is defined by
\beq
 G_{\mbox{{\scriptsize\bf pp}}'}(t)  =  \langle {\cal O}_{\mbox{{\scriptsize
{\bf p}}}} (t)
    {\cal O}_{\mbox{{\scriptsize {\bf p}}}'} (0)  \rangle  -
 \langle {\cal O}_{\mbox{{\scriptsize {\bf p}}}}(t) \rangle
 \langle {\cal O}_{\mbox{{\scriptsize {\bf p}}}'} (0) \rangle
\eeq
Inserting the complete set of states, one can rewrite the two-point function
as
\beq
 G_{\mbox{{\scriptsize\bf pp}}'}(t)=
\sum_{j} v_{\mbox{{\scriptsize {\bf p}}}}^{j}
v_{\mbox{{\scriptsize {\bf p}}} '}^j
       \ee^{-W_j t},
\eeq
with
$v^j_{\mbox{{\scriptsize {\bf p}}}}$ the coupling of the
state $j$ to the two-particle operator ${\cal O}_{\mbox{{\scriptsize\bf p}}}$.
The two-particle energy levels $W_j$'s can be
extracted by diagonalizing $G_{\mbox{{\scriptsize {\bf
p}}}\mbox{{\scriptsize {\bf p}}}'}(t)$ as a matrix in $\pbb,\pbb'$ at each time
slice $t$, and fitting the eigenvalues $\lambda_j(t)$ to a single
exponential\cite{LuescherWolff} \beq
    \lambda_j(t) \stackrel{t\rightarrow\infty}{\longrightarrow} C_j
       \ee^{-W_jt}.
\label{eq:eigen}
\eeq
In our calculation we truncate the momenta to the subset
$\mbox{{\bf p}}$ = (0,0,0), (1,0,0), (1,1,0), (1,1,1).  For
the lowest state a $\chi^2$ fit is made to the exponential form
(\ref{eq:eigen})
for $t=2-7$, while for higher states we used the range $t=2-4$ as
errors in the eigenvalues become quite significant for larger values of $t$.
The results for the two-particle energies are given in Table 2 where we also
list the
values of the momentum $k=\sqrt{(W/2)^2-m^2}$ with $m=E({\bf 0})$ taken from
Table 1.  The errors are estimated by the
jackknife method with the bin size of $10^6$ sweeps.

We have made a one-loop perturbative calculation of the phase shift to
compare with numerical results.  For the O($N$)-symmetric
$\phi^4$ theory in the continuum defined by the Lagrangian \beq
{\cal L}=\frac{1}{2}\sum_{\alpha=1}^{N}
         \del_{\mu}\phi^{\alpha} \del^{\mu}\phi^{\alpha}
          -\frac{1}{2} m_{\mbox{{\scriptsize R}}}^2 \sum_{\alpha=1}^{N}
            \phi^{\alpha} \phi^{\alpha}
          -\frac{1}{4!} g_{\mbox{{\scriptsize R}}} \left( \sum_{\alpha=1}^{N}
              \phi^{\alpha} \phi^{\alpha} \right) ^2
+ (\mbox{{\it counterterms}}),
\eeq
the two-particle scattering amplitude up to one-loop level,
renormalized by the momentum
subtraction at $p=0$, is given by
\beq
   T(\kb,\kb')=\frac{4\pi^2}{\omega_{k}^2} {\cal A}
   \left[ \alpha_{\mbox{{\scriptsize R}}} + \frac{1}{2}
   \alpha_{\mbox{{\scriptsize R}}}^2
      \{ {\cal A} \varphi(s)   +{\cal B} (\varphi(t) + \varphi(u) )
      + O(\alpha_{\mbox{{\scriptsize R}}}^3) \}  \right],
\eeq
where
$s$, $t$, $u$ are the Mandelstam variables constructed from the
initial and final momentum
$\kb$ and $\kb '$ in the center-of-mass frame,
$k\equiv |\kb|=|\kb'|$, $\alpha_{\mbox{{\scriptsize R}}}
=g_{\mbox{{\scriptsize R}}}/16 \pi ^2$,
and $\omega_{k} \equiv \sqrt{k^2+m_{\mbox{{\scriptsize R}}}^2}$.
The coefficients
take the values
${\cal A}=(N+2)/3$ and ${\cal B}=1$ for the scalar channel (
${\cal A}=2/3$ and ${\cal B}=(N+6)/6$ for the tensor channel),
and the function $\varphi$ is defined by
\beq
    \varphi(z) = \int _0 ^1 dx \mbox{ln} \left ( 1-
\frac{z}{m_{\mbox{{\scriptsize R}}}^2}
                   x (1-x) \right).
\eeq
The phase shifts $\delta_l (k)$ are defined through the partial-wave expansion
of $T(\kb,\kb')$
\beq
    T(\kb,\kb')=-\frac{8\pi}{\omega_{k}}\sum_{l=0}^{\infty}
    (2l+1) P_l (\cos \theta) \frac{\ee^{2i\delta_l(k)}-1}{2ik},
\eeq
with $\theta$ the angle between $\kb$ and $\kb'$.  The s-wave phase shift
$\delta_0 (k)$ is therefore obtained as
\beq
   \delta_0(k)=-\frac{\pi}{2}\frac{k}{\omega_k}
{\cal A}[\alpha_{\mbox{\scriptsize R}}+
\alpha_{\mbox{\scriptsize R}}^2 \{{\cal A}f(k)+{\cal B}g(k)\}
+ O(\alpha_{\mbox{\scriptsize R}}^3)],
\label{eq:perturbation}
\eeq
where the functions $f(k)$ and $g(k)$ are given by
\beqa
  f(k) &=& \frac{k}{\omega_{k}}\ln\frac{\omega_k+k}
  {m_{\mbox{{\scriptsize R}}}}-1,\\
 g(k) &=& \frac{1}{2} \int_0^{\pi}d\theta \sin\theta
 \left( \frac{\omega_{k\sin\frac{\theta}{2}}}
    {k\sin\frac{\theta}{2}}\ln\frac{\omega_{k\sin\frac{\theta}{2}}+
     k\sin\frac{\theta}{2}}{m_{\mbox{{\scriptsize R}}}}-1 \right)    \n
&~& \mbox{}+\frac{1}{2} \int_0^{\pi}d\theta \sin\theta
  \left( \frac{\omega_{k\cos\frac{\theta}{2}}}
    {k\cos\frac{\theta}{2}}\ln\frac{\omega_{k\cos\frac{\theta}{2}}+
     k\cos\frac{\theta}{2}}{m_{\mbox{{\scriptsize R}}}}-1 \right).
\eeqa

L\"{u}scher's formula (\ref{eq:luescher}) is derived in the continuum
space-time
and the dispersion relation of the one-particle energy $E({\bf
p})=\sqrt{{\bf p}^2+m^2}$ enters into the proof in an essential way.   The
simulation results for the one-particle energy are compared with the continuum
dispersion relation in fig.1. We find a good agreement
($E({\bf p})/\sqrt{{\bf p}^2+m^2}$ =0.994(6))
for the lowest momentum
$p=2\pi/L \sim 0.5$ for the lattice size $L=12$.  An increasing deviation for
larger momenta indicates   that finite lattice
spacing effects become non-negligible for higher momenta.

In fig.2 we show by filled circles our results for the s-wave phase shift
$\delta_0(k)$ in the scalar
channel extracted from the lowest two-particle state
(see Table 2 for numerical values). The dotted lines show the function $\phi
(kL/2\pi)$ and the solid line represents the one-loop perturbative result
(\ref{eq:perturbation}) calculated with the infinite volume estimates for the
renormalized parameters $m_{\mbox{{\scriptsize R}}}=0.3044$ and
$\alpha_{\mbox{{\scriptsize R}}}=0.142$ for
$\kappa=0.297$\cite{Frick}.   We have also used the data of ref.\cite{Frick}
for the lowest two-particle energy to calculate the phase shift.
The results are plotted by open circles in fig.2 and are consistent with
our values.  We observe a reasonable agreement between
the results of simulations and that of one-loop perturbation theory.   A trend
may be present, however, that the numerical results become smaller than the
perturbative prediction as the lattice size decreases.   This may represent a
systematic bias due to an   increase of vacuum polarization effects for small
lattice sizes, which are not taken into account in (\ref{eq:luescher}).

As is seen from the figure, the phase shift is almost linear in the region of
momenta which can be explored by the lowest two-particle states.
These states therefore do not give information on the phase shift
beyond the scattering length, which has already been calculated
\cite{Montvay,Frick} using the leading term of
L\"{u}scher's formula in $1/L$ expansion derived earlier\cite{Luescher2}.
In order to extract the phase shift for large momenta
we must therefore examine excited two-particle states.
The rightmost point in fig.3
shows the phase shift extracted from the first excited state for
$\kappa=0.297$ and $L=12$.
The agreement with the
perturbative prediction, albeit with a sizable error,
is quite encouraging.

We note that fig.3 reveals an important point in practical
applications of L\"{u}scher's formula. The function $\phi(kL/2\pi)$
increases very rapidly for momenta corresponding to excited states
(this is specific to four dimensions; in two dimensions
$\phi(kL/2\pi)=kL/2$ and the slope is constant\cite{LuescherWolff}).
As a consequence, a reliable extraction of the phase shift
for large momenta requires quite a precise determination of the
momentum $k=\sqrt{(W/2)^2-m^2}$, and hence that of the two-particle energy $W$.
For example, reducing the error of the phase shift for the first excited
state in fig.3 to a 10\% level necessitates a calculation
of the momentum
within 0.5\% accuracy compared to the 2\% error of the present data . Achieving
such an accuracy requires an order of magnitude  more computer time (our run
for the size $L=12$ took 20 hours for $4\times 10^6$ sweeps on HITAC S820/80
with the peak speed of 3 Gflops).  The requirement becomes rapidly more
demanding as we go to higher excited  states. For instance, with the present
statistics we only obtained  $-\delta_0(k)=0.4(9)$ for the second excited
state.

To conclude, the method works well for momenta which can be explored by the
lowest  two-particle state. We find, however, that extraction of phase shifts
for large momenta requires quite a high statistics determination
of excited two-particle state energies due to the steep increase of
the function $\phi(kL/2\pi)$.
We feel that this presents an obstacle in practical applications of the
method, with which  one can in principle extract scattering data beyond
scattering lengths in four dimensions.

{}~

{}~

I would like to thank Professor A.Ukawa for leading me to
this interesting subject and for continuous support.
I am also thankful to Professors M.Fukugita,
M.Okawa and Dr.H.Mino for kind advice.
The computations were carried out on HITAC S820/80 at KEK.

\newpage

\begin{table}
\centering
\begin{tabular}{ccccc}
\hline
${\bf p}$ & (0,0,0) & (1,0,0) & (1,1,0) & (1,1,1) \\
\hline
$L=8$ &  0.321(2) & 0.819(7) & 1.11(2) & 1.24(4) \\
$L=12$ & 0.305(2) & 0.602(3) & 0.783(6) & 0.91(1) \\
\hline
\end{tabular}
\caption{ Single-particle energy $E({\bf p})$ in lattice unit
for $\kappa=0.297$ for $L=8$ and
12.}
\vspace{5mm}
\label{table:disrel}
\end{table}

\begin{table}
\centering
\begin{tabular}{cccccc}
\hline
$L$ &$j$  & 0 & 1 & 2 & 3  \\
\hline
8 &$W_{j}$ & 0.73(2) & 1.6(2) & ---  &  ---   \\
   &$k_{j}$ & 0.17(2) & 0.74(11) & --- &  ---  \\
   &$-\delta_0(k_j)$ &  0.14(4) & --- & ---  &  ---  \\
 \hline
12 &$W_{j}$ & 0.646(9) & 1.26(3) & 1.66(9) & 1.72(15) \\
   &$k_{j}$ & 0.106(6) & 0.552(11) & 0.77(7) & 0.80(10) \\
   &$-\delta_0(k_j)$ & 0.12(2)  & 0.40(15)   &  0.4(9)  &  --- \\
 \hline
\end{tabular}
\vspace{2mm}
\caption{Two-particle energy $W_j$, the corresponding momentum
$k_j$ in lattice unit
and the s-wave phase shift $-\delta_0(k_j)$ calculated through (2) for
$\kappa=0.297$ for the size $L=8$ and 12.
Values for the cases when errors are too large are omitted.}
\label{table:twoenergy}
\end{table}

\bigskip

{}~

\bigskip

{}~

{}~

\newpage

\centerline{\Large Figure captions}
\bigskip
\noindent
Fig. 1
$E({\bf p})/\protect\sqrt{{\bf p}^2+m^2}$ as a function of $p=\vert{\bf
p}\vert$ for $L=8$(squares)
and
$L=12$ (circles)
at $\kappa=0.297$
with $m=0.321(2)$ for $L=8$ and $m=0.305(2)$ for $L=12$ as input.
\\

\bigskip
\noindent Fig. 2 S-wave phase shift calculated from the lowest two-particle
state  energy for $\kappa=0.297$ with various lattice sizes $L$.
Momentum $k$ is in lattice unit.
Dashed lines represent
the right hand side of (2) for each $L$.
Solid line represents the one-loop result(15).\\

\bigskip
\noindent  Fig. 3 S-wave phase shift calculated from the first
excited two-particle state
energy for $\kappa=0.297$ and $L=12$.
Those from the lowest ones are also plotted.\\

\end{document}